\documentclass[12pt,a4paper]{article}
\pdfoutput=1
\usepackage{color}
\usepackage{amssymb,amsmath,bm,bbm}
\usepackage{epsf}
\usepackage{epsfig}
\usepackage{afterpage}
\usepackage{longtable}
\usepackage[dvipsnames]{xcolor}
\usepackage[linktoc=page,bookmarks=false,colorlinks=false,
linkbordercolor=RoyalBlue,citebordercolor=ForestGreen,urlbordercolor=CornflowerBlue]{hyperref}
\usepackage{latexsym,mathrsfs,dsfont}
\usepackage[normalem]{ulem} 
\usepackage[compress]{cite}
\usepackage{graphicx}
\usepackage{url}
\usepackage{paralist}
\usepackage{bbold}

\newcommand{\ord}{\mathcal{O}}

\newcommand{\gev}{\, {\rm GeV}}

\newcommand{\bsi}{B_6^{(1/2)}}
\newcommand{\bei}{B_8^{(3/2)}}

\def\epe{\varepsilon'/\varepsilon}
\newcommand{\beq}{\begin{equation}}
\newcommand{\eeq}{\end{equation}}
\newcommand{\be}{\begin{equation}}
\newcommand{\ee}{\end{equation}}
\newcommand{\bi}{\begin{itemize}}
\newcommand{\ei}{\end{itemize}}
\newcommand{\ba}{\begin{array}}
\newcommand{\ea}{\end{array}}
\newcommand{\beqa}{\begin{eqnarray}}
\newcommand{\eeqa}{\end{eqnarray}}
\newcommand{\bea}{\begin{eqnarray}}
\newcommand{\eea}{\end{eqnarray}}
\newcommand{\beqn}{\begin{eqnarray}}
\newcommand{\eeqn}{\end{eqnarray}}

\definecolor{red}{cmyk}{0,1,1,0.4}

\def\Cut{\bf\color{red}}
\newcommand{\weak}{{\color{black}$\blacksquare$}}

\usepackage{fancyhdr}
\advance \headheight by 15.0truept       

\begin{document}


\vspace{-14mm}
\begin{flushright}
        {FLAVOUR(267104)-ERC-119}\\
CP3-16-08
\end{flushright}

\vspace{8mm}

\begin{center}
{\Large\bf
\boldmath{Final State Interactions in $K\to\pi\pi$ Decays:\\ $\Delta I=1/2$ Rule vs.  $\epe$ }}
\\[12mm]
{\bf \large  Andrzej~J.~Buras${}^a$ and Jean-Marc G\'erard${}^b$ \\[0.8cm]}
{\small
${}^a$TUM Institute for Advanced Study, Lichtenbergstr.~2a, D-85748 Garching, Germany\\
Physik Department, TU M\"unchen, James-Franck-Stra{\ss}e, D-85748 Garching, Germany\\[2mm]
${}^b$ Centre for Cosmology,
Particle Physics and Phenomenology (CP3), Universit{\'e} catholique de Louvain,
Chemin du Cyclotron 2,
B-1348 Louvain-la-Neuve, Belgium}
\end{center}

\vspace{4mm}

\begin{abstract}%
\noindent
Dispersive effects from strong $\pi\pi$ rescattering in the final state (FSI) of weak $K\to\pi\pi$ decays are revisited with the goal to have a global view on their
{\it relative} importance for the $\Delta I=1/2$ rule and the ratio $\epe$  
in the 
Standard Model (SM).  We point out that this goal cannot be reached within a pure effective (meson) field approach like {chiral perturbation theory}  in 
which the dominant current-current operators governing 
the $\Delta I=1/2$ rule and the dominant density-density (four-quark) 
operators governing $\epe$ cannot be disentangled from each other. 
But in the context of a dual QCD approach, which includes both long distance 
dynamics and the UV completion, that is QCD at short distance scales,
 such a distinction is possible. We find then that  beyond the strict large $N$ limit, $N$ being the number of colours, FSI 
are likely to be important for the $\Delta I=1/2$ 
rule but much less relevant for $\epe$. The latter finding diminishes significantly 
hopes that improved calculations of $\epe$
would 
 bring its SM prediction to agree with the experimental data,
 opening 
thereby an arena for important new physics contributions to this ratio.
\end{abstract}

\setcounter{page}{0}
\thispagestyle{empty}
\newpage


\section{Introduction}
Among  the most important observables in flavour physics are the ratio of $K\to\pi\pi$ isospin amplitudes ${\rm Re}A_0/{\rm Re}A_2$ and $\epe$. The first ratio
\be\label{N1a}
\frac{{\rm Re}A_0}{{\rm Re}A_2}=22.4\,,
\ee
expresses the so-called $\Delta I=1/2$ rule \cite{GellMann:1955jx,GellMann:1957wh} in $K\to\pi\pi$ decays.  On the other hand $\epe$ measured
by NA48 \cite{Batley:2002gn} and KTeV
\cite{AlaviHarati:2002ye,Abouzaid:2010ny} collaborations, to be
\be\label{EXP}
(\epe)_\text{exp}=(16.6\pm 2.3)\times 10^{-4} \,,
\ee
expresses CP-violation in $K\to\pi\pi$ decays.
 In the Standard Model (SM) the amplitudes 
 ${\rm Re}A_{0,2}$ are mostly governed by the  $Q_{1,2}$ current-current operators and $\epe$ by the QCD penguin $Q_6$ and electroweak penguin $Q_8$ density-density operators.
The most recent result for the $\Delta I=1/2$ rule from 
the dual approach to QCD reads \cite{Buras:2014maa}
\be\label{DRULEN}
\left(\frac{{\rm Re}A_0}{{\rm Re}A_2}\right)_{{\rm dual~QCD}}=16.0\pm1.5 \,,
\ee
while the corresponding result from the RBC-UKQCD collaboration is \cite{Bai:2015nea}
\be\label{DRULEL}
\left(\frac{{\rm Re}A_0}{{\rm Re}A_2}\right)_{{\rm lattice~QCD}}=31.0\pm11.1 \,.
\ee
Both results signal that ${\rm Re}A_0$ is strongly enhanced over
${\rm Re}A_2$ but there is a visible deficit in (\ref{DRULEN}) when compared with (\ref{N1a}), while the first lattice QCD 
result is still rather uncertain.

The present status of $\epe$ in the SM  can be summarized as follows. The
RBC-UKQCD lattice collaboration calculating hadronic matrix elements of 
all operators but not including isospin breaking effects finds 
\cite{Blum:2015ywa, Bai:2015nea}
\begin{align}
  \label{eq:epe:LATTICE}
  (\epe)_\text{SM} & = (1.38 \pm 6.90) \times 10^{-4},\qquad {\rm (RBC-UKQCD)}.
\end{align}
Using the hadronic matrix elements of QCD- and EW-penguin 
$(V-A)\otimes (V+A)$ operators from RBC-UKQCD lattice collaboration 
but extracting the matrix elements
of penguin $(V-A)\otimes (V-A)$ operators from the CP-conserving $K\to\pi\pi$ 
amplitudes and including isospin breaking effects one finds \cite{Buras:2015yba}
\begin{align}
  \label{eq:epe:LBGJJ}
  (\epe)_\text{SM} & = (1.9 \pm 4.5) \times 10^{-4},\qquad {\rm (BGJJ)}\,.
\end{align}
A new 
 result in \cite{Kitahara:2016nld} 
\begin{align}
  \label{KNT}
  (\epe)_\text{SM} & = (1.1 \pm 5.1) \times 10^{-4},\qquad {\rm (KNT)}\,.
\end{align}
confirms the findings in (\ref{eq:epe:LATTICE}) and (\ref{eq:epe:LBGJJ}) that 
the SM result for $\epe$ is significantly below its experimental value
in (\ref{EXP}). 

While these results, based on the hadronic matrix elements from RBC-UKQCD 
lattice collaboration, suggest some evidence for the presence of new physics (NP) in hadronic $K$ decays and favour NP models that are able to enhance $\epe$,
the large uncertainties in the hadronic matrix elements in question do not yet
preclude that eventually the SM will agree with data. In this context the 
upper bounds on the matrix elements of the
dominant penguin operators from large $N$ dual QCD approach 
\cite{Buras:2015xba} are important and  allow us to derive an upper bound on $\epe$
\begin{align}
  \label{BoundBGJJ}
  (\epe)_\text{SM} \le (8.6 \pm 3.2) \times 10^{-4}, \qquad {\rm (BG)}.
\end{align}
Moreover taking into account lattice results on the matrix elements of 
electroweak penguin operators ($\bei$) that are better known than those 
of QCD penguin operators ($\bsi$) one finds the values of $\epe$ significantly 
below this bound.

While the dual QCD approach allows to understand the suppression of $\epe$  in (\ref{eq:epe:LATTICE})-(\ref{KNT}) analytically, it does not yet properly 
include final state interactions (FSI). The question then arises whether these 
effects could improve the status of $\Delta I=1/2$ rule and of $\epe$ bringing the theory in both cases closer to data. In fact 
  the chiral perturbation theory (ChPT) practitioners, already long time ago, 
put forward the idea
that both the amplitude ${\rm Re}A_0$, governed by the current-current operator  $Q_2-Q_1$ and the $Q_6$ contribution to the ratio $\epe$ could be 
enhanced significantly through FSI in a correlated manner 
\cite{Antonelli:1995gw,Bertolini:1995tp,Frere:1991db,Pallante:1999qf,Pallante:2000hk,Buchler:2001np,Buchler:2001nm,Pallante:2001he}. The goal of this letter is to investigate whether this claim is really 
justified.

Before entering the details, let us make the following important observation that underlines the main points made in our paper.
The QCD penguin operator $Q_6$,  
generated by short-distance (SD) evolution from $M_W$ down to scales 
$\ord (1\gev)$ of the {\it current-current} four-quark operator $(Q_2-Q_1)$, is unambiguously identified as a {\it density-density }four-quark operator  
\cite{Shifman:1975tn,Buras:2014maa}. However such a distinction between $(Q_2-Q_1)$ and $Q_6$  is far from being evident during the further long-distance (LD) evolution below the critical 1~$\gev$  scale of QCD \cite{Fatelo:1994qh,Buras:2014maa}, though mandatory to consistently identify the strong FSI effects on the corresponding weak hadronic matrix elements.

\section{Weak hadronic matrix elements}

      In the standard ChPT approach based on the power counting in meson momenta, the weak $K$ decay amplitude for the dominant $\Delta I=1/2$ channel reads \cite{Cronin:1967jq}
\be\label{A0}
A_0=\langle \pi\pi(I=0)|\,G_8[\partial_\mu U\partial^\mu U^+]_{ds}\,|K\rangle\,,\qquad  {\rm at}~~~~\ord(p^2)
\ee
with $U(\pi)$, a unitary matrix transforming as $(3_L, 3_R^*)$ under global $U(3)_L\otimes U(3)_R$ transformations. Consequently, in this phenomenological approach the four-quark operators $(Q_2 - Q_1)$ and $Q_6$ contributing to $A_0$ are somehow merged into a single octet one, at least in the isospin limit \cite{Gerard:2005yk}. As a result, the corresponding current-current operator cannot be disentangled any more from the density-density operator. In the absence of any UV completion for this effective theory, their respective contributions to the $A_0$ decay amplitude (\ref{A0}) are encoded in the unique complex coupling $G_8$. Remarkably, this apparent merging of {\it a priori} quite different $\ord(p^2)$ operators can be seen at work once fundamental properties of QCD are eventually taken into account. 

       First of all, in the rather efficient large $N$ limit, $N$ being the number of colours\cite{'tHooft:1973jz,'tHooft:1974hx,Witten:1979kh}, both $\Delta S=1$ bosonized current-current \cite{Buras:1985yx} and density-density \cite{Bardeen:1986vp} operators factorize and reduce to form indeed the single octet operator given in (\ref{A0}). Fully exploiting the unitarity of the $U(\pi)$ matrix, one finds respectively
\be\label{Q12}
(Q_2-Q_1)\propto [\partial_\mu U U^+]_{dq}[\partial^\mu U U^+]_{qs}=-[\partial_\mu U\partial^\mu U^+]_{ds} \,,\qquad  {\rm at}~~~~\ord(p^2,0)
\ee
\be\label{Q6}
Q_6\propto [U-\frac{1}{\Lambda_\chi^2}\partial_\alpha\partial^\alpha U]_{dq}                
[U^+-\frac{1}{\Lambda_\chi^2}\partial_\beta\partial^\beta U^+]_{qs}= \frac{2}{\Lambda_\chi^2}[\partial_\mu U\partial^\mu U^+]_{ds} \,,\quad  {\rm at}~~~~\ord(p^2,0)
\ee
with $\Lambda_\chi$ a chiral breaking scale fixed by the $F_K/F_\pi$  ratio of pseudoscalar decay constants \cite{Chivukula:1986du,Bardeen:1986vp}
\be\label{Lchi}
\Lambda_\chi^2=F_\pi\frac{m_K^2-m_\pi^2}{F_K-F_\pi}\,.
\ee
 The ``$0$'' in 
$(p^2,0)$ indicates strict large $N$ limit: $1/N=0$.

      Secondly, in a dual QCD approach going beyond this strict large N factorization limit in a coherent way \cite{Buras:2014maa}, analytical tools allow us to keep distinguishing $(Q_2 - Q_1)$ from $Q_6$ operator even at the hadronic level through a matching of the slow SD quark-gluon evolution above 1~$\gev$ \cite{Bardeen:1986uz} with a fast LD meson evolution below 1~GeV \cite{Bardeen:1986vz}. Within such a dual frame based on a consistent $1/N$ expansion in the strong coupling $\alpha_s$ and $1/F_\pi^2$, the  hadronic matrix elements of the penguin operator $Q_6$ in question turn out to lie {\it below} its large $N$ value inferred from (\ref{Q6}) (and conventionally corresponding to $\bsi = 1$), namely \cite{Buras:2015xba}
\be\label{BG15}
\langle \pi\pi(I=0)|Q_6|K\rangle_{\rm dual\, QCD}\propto \bsi=1-\ord(\frac{1}{N})< 1\,,\quad  {\rm at}~~~\ord(p^2,0)+
\ord(p^0,1/N)\,.
\ee
Let us emphasize that the negative sign of the $1/N$ loop correction induced by the zero-derivative operator in (\ref{Q6}) is
 in agreement with the SD evolution of $\bsi$ parameter analyzed in \cite{Buras:1993dy}.
This then implies the result in (\ref{BoundBGJJ}), i.e.,
the $2\sigma$  tension when confronted with the measured CP-violating parameter \cite{Buras:2015xba,Buras:2015yba}. Yet, the question of $1/N$-suppressed strong FSI effects on such a hadronic matrix element may be raised at this point.

\begin{figure}[!tb]
 \centering
\includegraphics[width = 0.60\textwidth]{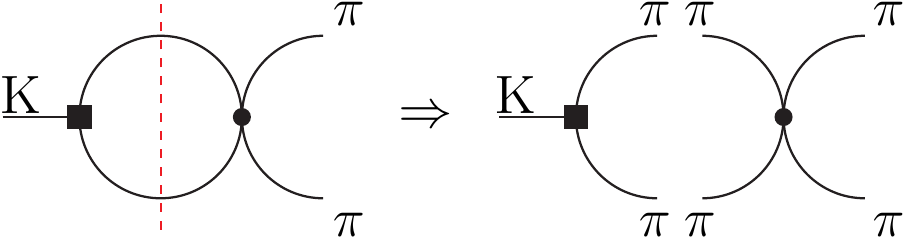}
\caption{Strong ($\bullet$) FSI effect on weak (\weak) hadronic 
matrix elements. The Cutcosky cut ({\Cut - - -}) tells us to put internal mesons on the mass-shell to consistently identify any $1/N$-suppressed absorptive part of the Feynman amplitude induced by the $Q_{1,2}$ and $Q_{6,8}$ operators.
}\label{fig:loop}
\end{figure}

\section{Strong final state interactions}
\subsection{Chiral perturbation theory and beyond}
In ChPT, strong phase shifts are zero in the leading-order approximation. In this analytical approach a pion loop  should be appended to any local weak $K\to\pi\pi$ transition in order to incorporate the strong $\pi\pi\to\pi\pi$ rescattering effects and, in particular,  non-zero FSI phase shifts.
Following the well-known Cutcosky cutting rule, this effective bubble triggers some ($1/N$-suppressed) absorptive part whenever the two mesons in the loop can be taken on-shell (see Fig.~\ref{fig:loop}). In the field theory, strong phases resulting from these final rescatterings are factorized in the corresponding isospin amplitudes, so that one can use the parametrization
\be\label{Watson}
A[K\to(\pi\pi)_I]\equiv A_I\exp(i\delta_I), \qquad (I=0,2)\,.
\ee 
In the limit of CP conservation, the amplitudes $A_I$ are real and positive by 
definition. They become complex quantities in the presence of CP violation. 
The measured $\delta_0$ angle being rather large compared to $\delta_2$, here one thus expects non-negligible higher-order dispersive corrections to $A_0$  since real and imaginary parts resulting from pion loops are necessarily linked by analyticity and unitarity constraints.

   Going now beyond ChPT (BChPT), one might then advocate \cite{Antonelli:1995gw,Bertolini:1995tp,Frere:1991db,Pallante:1999qf,Pallante:2000hk,Buchler:2001np,Buchler:2001nm,Pallante:2001he} that an overall dispersive factor $\mathcal{R}_0\approx \exp(1/N) > 1$ resulting from the all-order resummation of 
pion loops  only should be applied to the weak decay amplitude in (\ref{A0})
 and, in particular, to its indistinguishable penguin component. Doing such an exponential rescaling in the strict factorization limit (\ref{Q6})
for the $Q_6$ operator (i.e., for $\bsi=1$) to avoid any possible $1/N$ double counting, one would end up this time with a QCD penguin hadronic matrix element well {\it above} its large $N$ value  \cite{Antonelli:1995gw,Bertolini:1995tp,Frere:1991db,Pallante:1999qf,Pallante:2000hk,Buchler:2001np,Buchler:2001nm,Pallante:2001he},
\be\label{Pich}
\langle \pi\pi(I=0)|Q_6|K\rangle_{\rm BChPT}\propto \bsi \mathcal{R}_0 = 1+\ord(\frac{1}{N})>1\,,\quad  {\rm at}~~~\ord(p^2,0)+
\ord(p^2,1/N)\,.
\ee
This would imply a better agreement with the measured value of $\epe$ in
(\ref{EXP}) whenever the $\Delta I=1/2$ rule (\ref{N1a}) is assumed to begin with. However, resumming only part of higher order ChPT corrections into a simple dispersive factor is known to be dangerous. Moreover 
one should keep in mind that $\varepsilon^\prime$ in itself is proportional 
to the imaginary part of ratio $A_2/A_0$:
\be
\varepsilon^\prime=\frac{i}{\sqrt{2}}{\rm Im}\left(\frac{A_2}{A_0}\right)
\exp(i(\delta_2-\delta_0))\,.
\ee
Taken as such without using (\ref{N1a}), any overall increase of $A_0$ (and decrease of $A_2$) as proposed in \cite{Antonelli:1995gw,Bertolini:1995tp,Frere:1991db,Pallante:1999qf,Pallante:2000hk,Buchler:2001np,Buchler:2001nm,Pallante:2001he}  would then imply a decrease of $\epe$.

 But which bound on the $\bsi$ should one trust, the upper one (\ref{BG15}) from dual QCD or the lower one (\ref{Pich}) from chiral perturbation supplemented by a large N limit? 

\subsection{Dual QCD Approach}
       In the dual QCD approach \cite{Buras:2014maa},
the $\ord(p^2,1/N)$  bubble correction generated by the current-current
 operators $Q_{1,2}$ does also require some FSI dispersive rescaling. Indeed, the associated $\ord(p^2,0)$ on-shell tree-level amplitude corresponding to the right diagram in  Fig.~\ref{fig:loop} and
computed from (\ref{Q12}) is proportional to the SU(3)-breaking factor $(m^2_K - m^2_\pi)$ and thus non-vanishing.
However, the common $\ord(p^2,0)$ result in (\ref{Q12}) and (\ref{Q6}) of a strict large $N$ factorization does not necessarily imply that the $1/N$-suppressed FSI effects on $(Q_2-Q_1)$ and $Q_6$ matrix elements are identical.

 In fact, the (density-density) operator $Q_6$ can also generate the chiral octet operator  
$[\partial_\mu U\partial^\mu U^+]_{ds}$
through its zero-derivative   {term  $[U]_{dq}[U^+]_{qs}$} in (\ref{Q6}).
In general, the factorizable $1/N$ corrections to this term should exactly cancel the non-factorizable ones in order to preserve the unitarity of the $U(\pi)$ matrix. Yet, in our dual QCD approach these factorizable $1/N$ corrections to the bosonized $\bar q q$ densities are already included in the running of quark masses since the QCD mass terms $\bar q_L m q_R+h.c.$ are scale independent.  As a consequence, a non-zero $1/N$ contribution survives even after contracting the $q$ and $q^\prime$ flavour indices in the following non-factorizable LD evolution   \cite{Buras:2015xba}  from scale $\Lambda=\ord(1\gev)$ to scale $M=\ord(m_K)$

\be\label{eq:55}
U^{dq} U^{\dagger q^\prime s}(\Lambda)\rightarrow U^{dq} U^{\dagger q^\prime s}(M)- \frac{\ln(\Lambda^2/M^2)}{(4\pi F_\pi)^2}(\partial^\mu U\partial_\mu U^\dagger)^{ds}\delta^{q q^\prime}
.
\ee
This is the genuine $\ord(p^0,1/N)$  one-loop correction to the $Q_6$ hadronic matrix element \cite{Buras:2015xba}
\be\label{B6new}
\bsi = 1 -\frac{3}{2} \left(\frac{\Lambda^2_\chi}{(4\pi F_\pi)^2}\right)\ln(\frac{\Lambda^2}{M^2})\,,
\ee
 with $\Lambda_\chi^2(\approx 1 \gev^2)$, given in (\ref{Lchi}), a  sizeable momentum-independent substitute for $p^2(\approx m_K^2)$ as already outlined in   (\ref{BG15})  but obviously missing in (\ref{Pich}). Evidently this 
$\ord(p^0,1/N)$ contribution is absent in the matrix element of  the two-derivative operator $(Q_2-Q_1)$ in (\ref{Q12}) implying that $1/N$-suppressed loop effects on 
 $Q_6$ and $(Q_2-Q_1)$ matrix elements are not identical, in contrast to the claim 
made in \cite{Antonelli:1995gw,Bertolini:1995tp,Frere:1991db,Pallante:1999qf,Pallante:2000hk,Buchler:2001np,Buchler:2001nm,Pallante:2001he}.

In any analytical approach relying on some (truncated) expansion, what is called {\em FSI effects} might be a misnomer with respect to the well-defined Watson factorization theorem (\ref{Watson}) in field theory. In
this context one should carefully distinguish between {\it dispersive} and {\it absorptive} contributions from the $1/N$-suppressed loop diagrams in  Fig.~\ref{fig:loop}.

\begin{itemize}
\item
The operator $[U]_{dq}[U^+]_{qs}$ contributes to the left loop diagram with 
{\it off-shell} intermediate mesons and leads to the non-zero $\ord(p^0,1/N)$ {\it dispersive} term in (\ref{B6new}) calculated in  \cite{Buras:2015xba}.
This term competes with the $\ord(p^2,0)$ tree-level value of the $\bsi$  parameter normalized to one as possibly foreseen from a simultaneous expansion in 
$p^2 = \ord(\delta)$  and $1/N = \ord(\delta)$, the joint chiral and colour counting already invoked elsewhere \cite{Leutwyler:1996sa} for strong interaction physics. Being of the same order in $\delta$ as the leading term but having 
opposite sign, it is the main origin of the suppression of $\bsi$ and thus of $\epe$.
\item
Most importantly, following Cutcosky cutting rule, the operator $[U]_{dq}[U^+]_{qs}$ does not imply any {\it absorptive} part since, once again, the associated tree-level amplitude with on-shell pions in the right loop diagram of   Fig.~\ref{fig:loop} identically vanishes due to the unitarity property of the $U(\pi)$ matrix for the light pseudoscalars
\be\label{UNITARITY}
\langle\pi\pi(I=0)|[U]_{dq}[U^+]_{qs}| K\rangle_{\text{tree-level}} =0\, .
\ee
Consequently, the leading pion-loop contribution to the $Q_6$ matrix elements 
is purely dispersive such that $\bsi$ is under control. In contrast, the leading pion-loop contribution to the $Q_{1,2}$ matrix elements is {\em both} dispersive {\em and} absorptive.
\end{itemize}

This disparity between  the pion FSI effects on the matrix elements of $(Q_2-Q_1)$ and  $Q_6$ operators is the main result of our paper, which cannot be highlighted 
within an 
effective (meson) field approach like {chiral perturbation theory} where 
 these two operators are indistinguishable from the beginning.

The absence of on-shell rescattering impact on $\bsi$ at $\ord(p^0,1/N)$ gives us the confidence in the bound in  (\ref{BG15}) which is crucial for the suppression of $\epe$
below the data. This absence has been
 checked explicitly in \cite{Hambye:1998sma} through a full one-loop calculation of both factorizable and non-factorizable LD contributions that are generated by the density-density  penguin operator $Q_6$. In our dual QCD picture, the absorptive part of the former cannot be included in the running of quark masses while the absorptive part of the latter cannot be matched with SD evolution. So, they have to cancel each other, supporting in that manner the leading upper bound (\ref{BG15}) at the expense of the subleading lower bound (\ref{Pich}). 

    In fact, any attempt to include the first impact of strong FSI on $\bsi$ would require an expansion beyond the consistent $\ord(\delta)$  bound (\ref{BG15}). Unfortunately a full $\ord(\delta^2)$ estimate of the $Q_6$  matrix element, with further $\ord(p^2,1/N)$ as well as genuine $\ord(p^4,0)$ and $\ord(p^0,1/N^2)$ corrections in (\ref{Pich}), is a task beyond the authors present skills. 
At best, we can quote the following {\it partial} results  

\be
\delta\bsi(p^4,0) \supset \frac{(m_K^2+m_\pi^2)}{2\Lambda_\chi^2}\approx +0.15
\ee
from (\ref{Q6}) alone  and 
\be
\delta\bsi(p^0,1/N^2) \supset -\frac{4}{N}\left(\frac{m_0}{4\pi F_\pi}\right)^2\approx -0.35
\ee
from the anomalous effective Lagrangian that solves the so-called $U(1)_A$ problem \cite{Buras:2015xba}. These versatile numbers encourage us to stick to a consistent $\ord(\delta)$ calculation for $\bsi$ rather than to venture in an unreliable  $\ord(\delta^2)$  estimate of this hadronic parameter. In other words, our
upper bound (\ref{BG15}) on $\bsi$ follows from the above $\delta$ expansion under the assumption 
\be
\ord(\delta^2)<\ord(\delta)\,,
\ee
while the lower bound (\ref{Pich}) would require large $\ord(\delta^2)$ corrections to be true. After all, 
 the pseudoscalar mass spectrum is reproduced within $15\%$ on the sole basis of the effective Lagrangian for strong interactions at $\ord(\delta)$ \cite{Gerard:2004gx}, with the axial $U(1)$ breaking scale $m_0=\ord(0.85\gev)$ associated to a large $\eta^\prime$ mass, its $\ord(p^0,1/N)$ component.

\section{Comments and conclusion}

    The dispersive rescaling factors $R_0\approx 1.55$ and $R_2\approx 0.92$, 
corresponding respectively to $\delta_0\gg 0$ and $\delta_2<0$, have been extracted from an all-order resummation of the 1/N-suppressed FSI in \cite{Pallante:2000hk,Pallante:2001he}. Naively applied to the hadronic matrix elements of the free $|\Delta S| = 1$ weak Hamiltonian to avoid, once again, any possible double counting, they would imply the following $\Delta I = 1/2$ enhancement:
\be\label{Pich1}
\frac{{\rm Re}A_0}{{\rm Re}A_2}= \sqrt{2}\times \frac{\mathcal{R}_0}{\mathcal{R}_2}\approx 2.4\,.
\ee
In this rather peculiar large $N$ limit indeed, only the $Q_2$ operator with its two charged currents survives such that the $K^0\to\pi^0\pi^0$ neutral channel is purely induced by $\pi^+\pi^-\to\pi^0\pi^0$ rescattering.
Further $\ord(1/N)$ corrections from strong interactions,  namely LD and SD evolutions \cite{Buras:2014maa}, are obviously required to understand the measured value in (\ref{N1a}).
Whatever the approach adopted these corrections must also be large, even if formally $\ord(p^2, 1/N)$, and properly combined with the FSI LD one given in (\ref{Pich1}). 

    Similarly, the FSI rescaling factors $\mathcal{R}_{0,2}$ applied to a strict large $N$ value of the QCD and electroweak penguin hadronic matrix elements (i.e., the one obtained for $\bsi = 1$ and $\bei=1$), namely
\be
\bsi=1\times \mathcal{R}_0\approx 1.55\,, \qquad \bei=1\times \mathcal{R}_2\approx 0.92\
\ee
would also miss strong $\ord(p^0)$ and mild $\ord(p^2)$ $1/N$  contributions, respectively. Again, such a disparity between FSI effects on $\bsi$ and $\bei$ is due to the fact that
\be
\langle\pi\pi(I=2)|[U]_{dq}e_q[U^+]_{qs}| K\rangle_{\text{tree-level}}  \not=0
\ee
instead of (\ref{UNITARITY}) when the quark electric charges $e_q$ are introduced.

 Relying now more specifically on a simultaneous expansion in $p^2 = \ord(\delta)$ and $1/N =\ord(\delta)$ in the dual QCD approach involving both SD and LD operator evolutions at the one-loop level, we come then to the following conclusions.
\begin{itemize}
\item
The all-order resummation of FSI effects from the $Q_{1,2}$ current-current operators would definitely help filling the persistent gap of about $30\%$ between theory and experiment for the $\Delta I = 1/2$  rule  \cite{Buras:2014maa},
 though some $\ord(p^2,1/N)$ double counting at the LD level seems difficult to avoid within present analytical techniques relying on some expansion. Here 
non-perturbative approaches like lattice QCD  could turn out to be more
successful. In lattice computations, the strong phases are determined using the Luscher relation between the two-pion energies in a finite volume and the phase-shifts \cite{Luscher:1985dn,Luscher:1986pf}. The moduli are fully calculated \cite{Luscher:1985dn,Luscher:1986pf,Luscher:1990ux,Luscher:1991cf,Lellouch:2000pv} and the amplitudes are then given by (\ref{Watson}).
\item
 The first FSI effects induced by the $Q_6$ density-density operator being subleading in either $p^2$ or $1/N$ within an appropriate chiral/color expansion,  they do not really relax the tension recently highlighted in \cite{Buras:2015xba,Buras:2015yba} for the CP-violating parameter $\epe$.
\end{itemize}

     In other words, the FSI rescaling factors $ \mathcal{R}_I$ extracted from dispersive treatments beyond one-loop  \cite{Antonelli:1995gw,Bertolini:1995tp,Frere:1991db,Pallante:1999qf,Pallante:2000hk,Buchler:2001np,Buchler:2001nm,Pallante:2001he}  are relevant for the $\Delta I = 1/2$ rule in  \cite{Buras:2014maa}, enhancing the expectations that the $\Delta I=1/2$
rule is fully governed by SM dynamics. 

On the other hand our findings imply that FSI are much less relevant for 
 $\epe$ and diminish  significantly 
hopes that improved  calculations of $\epe$
would 
 bring it within the SM to agree with the experimental data, opening 
thereby an arena for important new physics contributions to this ratio. For latest analyses 
of such contributions see \cite{Blanke:2015wba,Buras:2015yca,Buras:2015kwd,Buras:2015jaq,Buras:2016dxz,Tanimoto:2016yfy,Kitahara:2016otd,Endo:2016aws,Cirigliano:2016yhc}.

\section*{Acknowledgements}
We would like to thank Robert Buras-Schnell for comments on the manuscript and 
the construction of Fig.~\ref{fig:loop}. The discussions on FSI effects with Chris Sachrajda 
are appreciated. 
This research was done and financed in the context of the ERC Advanced Grant project ``FLAVOUR''(267104) and the Belgian IAP Program BELSPO P7/37. It was also
 partially
supported by the DFG cluster
of excellence ``Origin and Structure of the Universe''.

\renewcommand{\refname}{R\lowercase{eferences}}

\addcontentsline{toc}{section}{References}

\bibliographystyle{JHEP}
\bibliography{BG16}
\end{document}